# Giant enhancement of interlayer exchange in an ultrathin 2D magnet

Dahlia R. Klein[1†], David MacNeill[1†], Qian Song[2], Daniel T. Larson[3], Shiang Fang[3], Mingyu Xu[4,5], R. A. Ribeiro[4,5,6], P. C. Canfield[4,5], Efthimios Kaxiras[3,7], Riccardo Comin[1], Pablo Jarillo-Herrero[1*]

[1]Department of Physics, Massachusetts Institute of Technology, Cambridge, MA 02139, USA
[2]Department of Materials Science & Engineering, Massachusetts Institute of Technology, Cambridge, MA 02139, USA
[3]Department of Physics, Harvard University, Cambridge, MA 02138, USA
[4]Ames Laboratory, U. S. Department of Energy, Iowa State University, Ames, IA 50011, USA
[5]Department of Physics and Astronomy, Iowa State University, Ames, IA 50011, USA
[6]Centro de Ciências Naturais e Humanas, Universidade Federal do ABC, Santo André, SP, Brazil
[7]John A. Paulson School of Engineering and Applied Sciences, Harvard University, Cambridge, MA 02138, USA

[†]These authors contributed equally to this work.
[*]Correspondence to: pjarillo@mit.edu

**Main Text:**

Following the recent isolation of monolayer $CrI_3$[1], there has been a surge of new two-dimensional van der Waals magnetic materials[2-12], whose incorporation in van der Waals heterostructures offers a new platform for spintronics[5-8], proximity magnetism[13], and quantum spin liquids[14]. A primary question in this burgeoning field is how exfoliating crystals to the few-layer limit influences their magnetism. Studies on $CrI_3$ have shown a different magnetic ground state for ultrathin exfoliated films[1,5,6] but the origin is not yet understood. Here, we use electron tunneling through few-layer crystals of the layered antiferromagnetic insulator $CrCl_3$ to probe its magnetic order, finding a ten-fold enhancement in the interlayer exchange compared to bulk crystals. Moreover, temperature- and polarization-dependent Raman spectroscopy reveal that the crystallographic phase transition of bulk crystals does not occur in exfoliated films. This results in a different low temperature stacking order and, we hypothesize, increased interlayer exchange. Our study provides new insight into the connection between stacking order and interlayer interactions in novel two-dimensional magnets, which may be relevant for correlating stacking faults and mechanical deformations with the magnetic ground states of other more exotic layered magnets, such as $RuCl_3$[14].

A key family of van der Waals magnets is the layered transition metal trihalides, which have been studied for decades as prototypical magnetic insulators[15-17] and as a platform for quasi-two-dimensional magnetism[18-20]. In the chromium trihalides, the Cr atoms are arranged in a honeycomb structure, with each Cr atom surrounded by six halide atoms in an octahedral geometry (Fig. 1a). The bulk crystals undergo a crystallographic phase transition from a monoclinic phase (space group $C2/m$) at room temperature to a rhombohedral phase (space group $R\bar{3}$) at low temperatures (below about 240 K for $CrCl_3$[21]). While the intralayer lattice spacings are largely unaffected by this transition, the layer stacking sequence changes dramatically from rhombohedral ABC ordering at low temperatures to $AA_{1/3}$ stacking above 240 K, where each layer is displaced along the *a* axis by 1/3 of a lattice vector (Fig. 1a). Bulk $CrI_3$ and $CrBr_3$ are ferromagnetic below their Curie temperatures with moments aligned perpendicular to the *ab* plane; in contrast, bulk $CrCl_3$ is an antiferromagnet below its Néel temperature of 14 K[21-23]. The Cr moments in each layer of $CrCl_3$ are ferromagnetically coupled, but couple antiferromagnetically across the van der Waals gap (Fig. 1b). However, this interlayer exchange coupling is weak and the magnetization of adjacent layers in bulk $CrCl_3$ can be aligned with a small in-plane magnetic field of about 0.2-0.25 Tesla[21,24].

The nature of stacking in few-layer chromium trihalides has recently garnered attention owing to the intriguing observation of an antiferromagnetic (AFM) ground state and interlayer exchange in few-layer $CrI_3$, in contrast to the ferromagnetic (FM) ground state of bulk crystals [1,3,5,6,25,26]. This crossover from FM to AFM interlayer coupling has led to theoretical investigations centered on the origin of the antiferromagnetic ground state in bilayer $CrI_3$[27-29]. First principles calculations predict that while the rhombohedral phase favors FM interlayer exchange, the monoclinic phase prefers an AFM alignment between the layers[29]. Thus, it is natural to propose that the observed AFM coupling results from a monoclinic stacking that persists to low temperatures in ultrathin $CrI_3$. However, this proposal has not been verified experimentally. For $CrCl_3$, both the crystal and magnetic structures of ultrathin films are unknown, and could provide valuable insight into how stacking order and exfoliation couple to interlayer interactions in these materials.

In this work, we use electron tunneling to probe the magnetic structure and interlayer exchange coupling of few-layer $CrCl_3$ crystals. We first use mechanical exfoliation to obtain crystallites ranging from two to four layers in thickness. We then fabricate vertical magnetic tunnel junctions using two few-layer graphite electrodes (see Methods). We assemble the stacks using few-layer graphite flakes above and below each $CrCl_3$ flake and encapsulate the entire van der Waals heterostructure in top and bottom hexagonal boron nitride (Fig. 1c). In the final step, each stack is transferred onto a silicon substrate with a 285 nm oxide layer also containing prepatterned Ti/Pd wires; the stack is aligned so that the two few-layer graphite electrodes contact the Ti/Pd wires. The junctions show a high resistance (Fig. 1d and Supplementary Materials) with a nonlinear current-voltage relationship characteristic of tunneling in the Fowler-Nordheim regime [30].

We find that the junction conductance increases dramatically when a magnetic field is applied in the plane of the crystal layers. Fig. 2a shows the differential conductance for a tetralayer device as a function of bias, at 300 mK and in-plane applied fields of 0 and 2 Tesla, where the AC excitation is 50 mV. The junction differential magnetoresistance

$$\text{MR} = \frac{\frac{dI}{dV}(\mu_0 H_\parallel) - \frac{dI}{dV}(\mu_0 H_\parallel = 0)}{\frac{dI}{dV}(\mu_0 H_\parallel = 0)} \times 100\% \qquad (1)$$

is plotted in Fig. 2b, where $\mu_0 H_\parallel$ is 2 Tesla for the high-field measurement. Consistent with previous results[6] and the theory of spin filter tunneling[31], the differential magnetoresistance peaks at a finite bias value near the onset of Fowler-Nordheim tunneling (between 625 and 875 mV). This peak corresponds to a maximum of spin-polarized tunneling when the applied bias is sufficient to allow efficient Fowler-Nordheim tunneling for majority-spin electrons, which experience a lower energy barrier in the CrCl$_3$ when it is polarized, but not for minority-spin electrons, which experience a larger energy barrier. We further investigate the temperature and field dependence of this phenomenon by plotting the tunneling conductance of a CrCl$_3$ bilayer device as a function of both magnetic field and temperature (Fig. 2c). A vertical line cut of conductance versus temperature at zero applied field (Fig. 2d) shows a marked decrease in the conductance between 14 and 16 K. This drop reflects the onset of AFM alignment between the two layers in the junction, in agreement with the bulk Néel temperature[21]. If the interlayer exchange were instead FM, one would expect increased conductance in the magnetic state due to the lowering of the energy barrier for majority-spin electrons[32].

In principle, there are two mechanisms that can provide magnetoresistance in our junctions: increasing the saturation magnetization *within* each layer (which decreases the majority-spin barrier[33]), and rotation of the magnetization in *adjacent* layers from antiparallel to parallel as the external field is increased (called the double spin filter effect[31,34]). We expect that the latter effect is dominant in our junctions because: i) the observed Néel temperature is similar to bulk crystals in which case the magnetization in each layer is nearly saturated at a value of 3 Bohr magnetons per Cr atom by 10 K, and ii) we observe only a single kink in the conductance versus magnetic field curves, rather than the two that would be expected for an alignment of the layers followed by a saturation of the intralayer magnetization.

To analyze our conductance versus applied magnetic field data at 4.2 K and to extract the interlayer exchange, we model the magnetization as fully saturated within each layer and coupled between layers via an AFM exchange field $H_E$ (see Supplementary Materials for the precise definition of our model and interlayer exchange field). The experimental magnetoresistance curves are shown in Fig. 3a. Consistent with the spin filter effect when tunneling through a magnetic insulating barrier[34,35], the magnetoresistance increases with barrier thickness from 9% in a bilayer CrCl$_3$ device up to 208% in a tetralayer CrCl$_3$ device (Fig. 3c). The normalized magnetoresistance curves are displayed in Fig. 3b to emphasize the difference in saturation fields. Surprisingly, in all devices, the AFM to FM transition occurs at a much higher field than for bulk crystals (roughly 0.2-0.25 Tesla[21,24]). These AFM-FM transition fields range from 0.85 Tesla in the bilayer junction to 1.65 Tesla in the tetralayer junction. The reported bulk interlayer exchange field is 0.084 Tesla[18]. However, based on the saturation fields in Fig. 3a, we calculate increased values for the exchange coupling of 0.86 Tesla, 0.96 Tesla, and 0.97 Tesla in bilayer, trilayer, and tetralayer CrCl$_3$, respectively (see Supplementary Materials for calculation of the interlayer exchange from the experimentally observed metamagnetic transition). The consistency of the extracted interlayer exchange coupling for different thicknesses, despite large differences in the saturation field, strongly supports our hypothesis that the magnetoresistance arises from alignment of magnetic moments in adjacent layers. These magnetoresistances thus demonstrate more than a ten-fold increase in the interlayer exchange strength when CrCl$_3$ is cleaved to the few-layer limit. Furthermore, the thickness dependence of the interlayer exchange in the few-layer films is weak and the interlayer exchange is seen to slightly *increase* with thickness in our devices. This suggests that the dramatic enhancement is due to a qualitatively different property of the exfoliated films compared to bulk (*i.e.* their stacking order).

In order to explore our hypothesis of stacking order as the origin of enhanced interlayer exchange, we performed Raman spectroscopy on both bulk and exfoliated thin CrCl$_3$ crystals (Fig. 4). The experiments were performed in a backscattering geometry with a 532 nm laser incident perpendicular to the crystal *ab* plane. The incident light was linearly polarized and inelastically-scattered photons were detected in the parallel-polarized channel (XX). The 247 cm$^{-1}$ Raman mode of bulk CrCl$_3$ undergoes a marked peak shift close to 2 cm$^{-1}$ at the crystallographic phase transition near 240 K[36]. This excitation corresponds to out-of-plane vibrations of Cl atoms in the *ab* plane, which depend sensitively on the shearing of layers with respect to one another. To calibrate the peak position, we also fit the 209 cm$^{-1}$ Raman mode, which does not shift at the phase transition, and plot the difference between the two peak energies. We first studied the evolution of this energy difference upon cooling of a bulk crystal (Fig. 4a). Consistent with previous work[36], we observe a sharp shift in the peak energy difference between 240 and 250 K (see Supplementary Materials for complete spectra). Next, we examined the same peaks in an exfoliated crystal (35 nm thick) on a SiO$_2$/Si substrate. The peak difference shows minimal temperature dependence and no evidence of a transition even down to 10 K (Fig. 4b). We find similar results in an even thinner (8 nm thick) exfoliated crystal, with no apparent phase transition down to 180 K (Fig. 4c). This absence of a peak shift suggests that thin exfoliated CrCl$_3$ crystals remain in the high temperature monoclinic phase at low temperatures. Another explanation is that deformation of the crystal during the aggressive exfoliation process introduces stacking faults. In fact, repeated deformation of a bulk *α*-RuCl$_3$ crystal by bending back and forth gives rise to stacking faults leading to a change in the Néel temperature[37]. Thus, the exfoliation process itself likely plays a role in modifying the physical properties of thin van der Waals crystals compared to their original bulk structures.

To further study the crystal phases of exfoliated versus bulk CrCl$_3$, we analyzed the energy of the 247 cm$^{-1}$ Raman peak as a function of polarization angle. In the rhombohedral $R\bar{3}$ phase this peak is produced by a doubly-degenerate $E_g$ mode, whereas in the monoclinic *C*2/*m* phase it arises from a combination of closely spaced $A_g$ and $B_g$ modes[38,39]. In both cases, the two modes contributing to the observed peak have opposite polarization dependence. When they are degenerate, the opposite polarization dependences of these modes cancel out, leading to a single peak with constant intensity at one energy, as seen in the $R\bar{3}$ phase. But when the two modes are slightly offset in energy, their out-of-phase intensity variations lead to the observation of a combined peak whose peak position displays a fourfold modulation as a function of polarization angle (see Supplementary Materials for a detailed analysis). Based on these symmetry arguments, we can determine the crystal structure of bulk and thin films via polarized Raman spectroscopy. For a bulk crystal at 300 K (and therefore in the *C*2/*m* phase), we observe a fourfold variation in the energy of this peak as we sweep the polarization angle (Fig. 4d). However, upon cooling the bulk sample to 80 K, the peak energy oscillations become an order of magnitude smaller. This change is indicative of the crystallographic phase transition to the $R\bar{3}$ phase by 80 K (the small, resolution-limited residual signal may arise from some regions pinned in the high-temperature structure, see Ref. 22). In contrast, we found that a thin (17 nm thick) exfoliated flake on a SiO$_2$/Si substrate has strong fourfold variation in the 247 cm$^{-1}$ mode peak position at both 300 K and 80 K (Fig. 4e). This result gives strong additional evidence that exfoliated CrCl$_3$ is in the monoclinic phase even at lower temperatures.

We note that the oscillation of the 247 cm$^{-1}$ Raman peak energy in the exfoliated crystal is somewhat smaller at 80 K than 300 K. This could arise from either a partial transition of the crystal into the low temperature phase, or some inherent temperature dependence of the peak oscillation due to inhomogeneous hardening of the closely-lying phonon lines. We cannot rule out either scenario, but expect that our junctions are uniformly in the monoclinic stacking order. This is because we only observe a single kink in the magnetoresistance versus field traces, ruling out the presence of a highly nonuniform interlayer exchange coupling. In one trilayer device (see

Supplementary Materials) we did see a sharp increase in the magnetoresistance saturating around 0.15 Tesla, which could arise from a stacking fault within the barrier creating coexisting regions of monoclinic and rhombohedral stacking.

To understand the dependence of the interlayer exchange on the stacking type, we carried out density functional theory (DFT) calculations of the energy difference between the FM and AFM states for bilayer $CrCl_3$ with various stacking arrangements (see Supplementary Materials). We find that the *C*2/*m*-type stacking in the bilayer strongly favors an AFM interlayer alignment whereas the $R\bar{3}$-type phase can favor AFM or FM coupling, depending on the chosen DFT functional and interlayer spacing. Furthermore, the *difference* in interlayer exchange energy between the *C*2/*m*-type and $R\bar{3}$-type structures is relatively insensitive to the DFT functional used, and shows that the *C*2/*m*-type phase always has significantly larger AFM coupling than the $R\bar{3}$-type phase. For a realistic range of interlayer spacings from 0.55 nm to 0.65 nm, the calculated value of the interlayer exchange difference is on the same order as our experimental estimate. Overall, our calculations show that the *C*2/*m*-type phase has stronger AFM interlayer coupling than the $R\bar{3}$-type phase, consistent with the observed giant enhancement of interlayer exchange in ultrathin *C*2/*m* $CrCl_3$.

In summary we have demonstrated a ten-fold increase in the interlayer exchange of ultrathin $CrCl_3$ crystals. Employing spin-filter magnetic tunnel junctions fabricated from few-layer $CrCl_3$ barriers, we have determined that the magnetic order in individual sheets of the crystal couple more strongly across the van der Waals gap, resulting in a much higher in-plane saturation field. Furthermore, using temperature- and polarization-dependent Raman spectroscopy, we attribute this dramatic enhancement to a change in the stacking order of thin films compared to bulk crystals. This result has important implications for understanding the unexpected magnetic behavior in the ultrathin chromium trihalides, and shows that stacking order and faults must be considered in any experiment involving exfoliated halide magnets. For example, the FM to AFM transition observed in ultrathin $CrI_3$ is likely also a result of *C*2/*m* stacking in the few-layer crystals, as has been predicted from first principles calculations[29]. Moreover, our study opens the possibility to drastically modify exchange couplings in other van der Waals magnets by manipulating the stacking order between the layers through strain, twisting, and other methods[40-43].

**Acknowledgements:** This work was supported by the Center for Integrated Quantum Materials under NSF Grant DMR-1231319 (D.R.K., E.K., and S.F.), the DOE Office of Science, Basic Energy Sciences under award DE-SC0018935 (D.M.), as well as the Gordon and Betty Moore Foundation's EPiQS Initiative through grant GBMF4541 to P.J.-H. D.R.K. acknowledges partial support by the NSF Graduate Research Fellowship Program under Grant No. 1122374. R.C. acknowledges support from the Alfred P. Sloan Foundation. X.S. is supported by the Xu Xin International Student Exchange Scholarship from Nanjing University. E.K. and S.F. are also supported by the ARO MURI Award No. W911NF-14-0247. Work done at Ames Laboratory (M.X., R.A.R., and P.C.) was performed under Contract No. DE-AC02-07CH11358. R.A.R. was supported by the Gordon and Betty Moore Foundation's EPiQS Initiative through Grant GBMF4411. The computations in this paper were run on the Odyssey cluster supported by the FAS Division of Science, Research Computing Group at Harvard University.

**Author Contributions:** D.R.K., D.M., and P.J.-H. conceived the project. D.R.K. and D.M. grew the bulk $CrCl_3$ crystals, fabricated and measured the transport devices, and analyzed the data. Q.S. carried out Raman measurements under supervision of R.C. D.T.L. and S.F. carried out symmetry analysis and DFT calculations under supervision of E.K. M.X., R.A.R., and P.C. supplied the boron nitride crystals. All authors contributed to writing the manuscript.

**Competing Financial Interests:** The authors declare no competing financial interests.

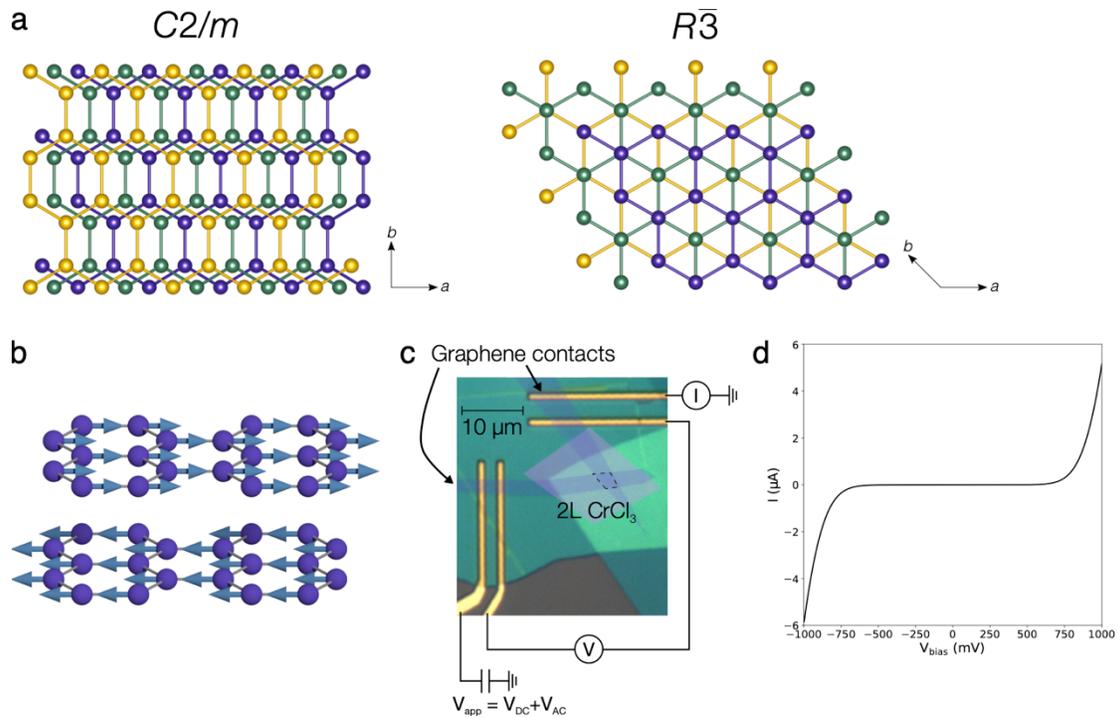

**Figure 1 | (a)** Stacking sequence of the Cr honeycomb sublattice in the *C*2/*m* (high temperature, left) and $R\bar{3}$ (low temperature, right) phases of CrCl$_3$ viewed perpendicular to the *ab*-plane. In both cases, the first and fourth layers are approximately aligned. **(b)** Schematic of moments in two adjacent layers at equilibrium. The moments lie in the *ab* plane. **(c)** False-color optical microscope image of a bilayer CrCl$_3$ tunnel junction device with four-point contact geometry for differential conductance measurements. **(d)** Current versus bias voltage for a trilayer tunnel junction at 4.2 K. The current was obtained by integration of the differential conductance with an AC bias voltage excitation of 50 mV.

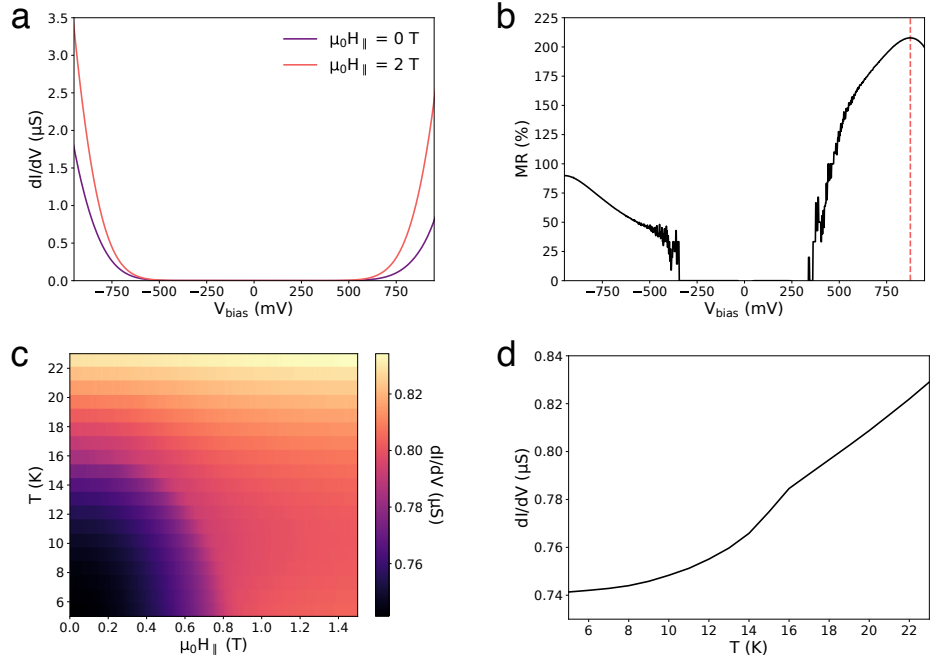

**Figure 2 | (a)** Differential conductance versus bias voltage for a tetralayer tunnel junction with an AC excitation of 50 mV at 300 mK. The data are taken at zero applied field (pink) and with an in-plane magnetic field of 2 Tesla (purple). **(b)** Magnetoresistance percent versus applied bias voltage for a tetralayer tunnel junction with a high-field in-plane magnetic field of 2 Tesla, extracted from **(a)**. The dashed line at a bias of 875 mV indicates the optimum bias for magnetoresistance. **(c)** Differential conductance versus applied in-plane magnetic field and temperature for a bilayer tunnel junction. The DC bias is -625 mV and the AC excitation is 50 mV. **(d)** Vertical line cut from **(c)** at zero applied magnetic field. The kink between 14 K and 16 K indicates the Néel temperature.

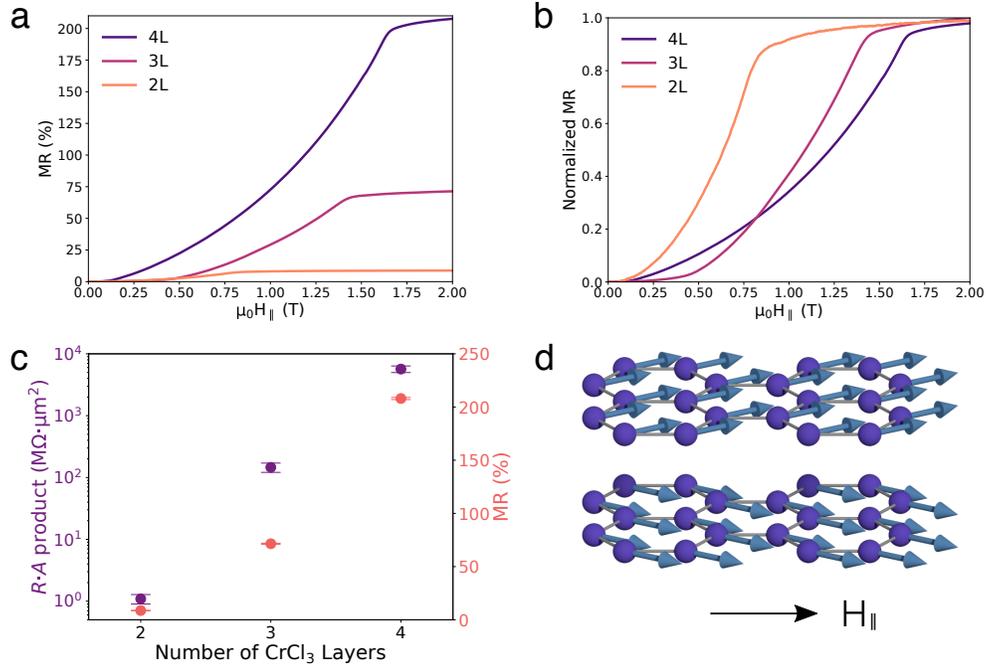

**Figure 3 | (a)** Magnetoresistance versus applied in-plane magnetic field for bilayer, trilayer, and tetralayer tunnel junctions at finite applied bias and an AC excitation of 50 mV at 4 K. **(b)** Normalized magnetoresistance versus applied in-plane magnetic field. **(c)** Differential resistance-area product (purple) and magnetoresistance (pink) versus $CrCl_3$ layer number. The differential resistance was measured with an applied DC bias of 500 mV at 4 K for each device. **(d)** Schematic of moments in two layers canting towards the in-plane applied field.

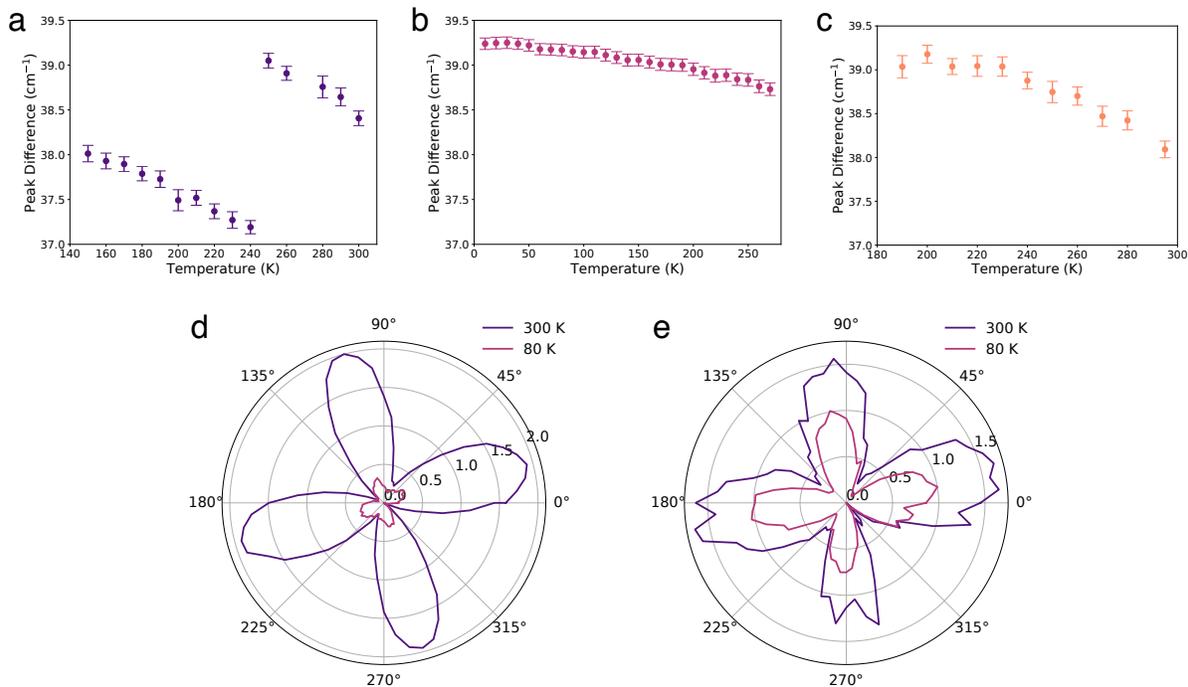

**Figure 4 | (a)** Difference between Raman peak positions of the 247 cm$^{-1}$ and 209 cm$^{-1}$ modes versus temperature for a bulk crystal of CrCl$_3$. A jump in the peak difference occurs around the crystallographic phase transition near 240 K. **(b)** The same peak difference versus temperature for an exfoliated 35 nm thick flake of CrCl$_3$ on a 90 nm SiO$_2$/Si substrate. The peak difference smoothly evolves down to 10 K without evidence of a phase transition. **(c)** The same peak difference versus temperature for an exfoliated 8 nm flake. **(d)** Peak position shift (in cm$^{-1}$) relative to minimum of 247 cm$^{-1}$ mode versus polarization angle for a bulk crystal. **(e)** Same as **(d)** for an exfoliated 17 nm thick flake on a 90 nm SiO$_2$/Si substrate.

## Methods:

### Bulk crystal growth

Bulk $CrCl_3$ crystals were grown by recrystallization of anhydrous $CrCl_3$ flakes. Approximately 1.0 g of the flakes (99.9%, Alfa Aesar) was loaded into a silica ampule in an inert environment and sealed under vacuum. The ampule was placed in a three-zone tube furnace with source, growth, and third zones held at 700ºC, 550ºC, and 625ºC, respectively, for a duration of 6 days. The source material was fully transported to the middle growth zone where it recrystallized into flat platelet crystals.

Hexagonal boron nitride (hBN) single crystals were grown from a high temperature, high pressure solution with an atomic Mg:B ratio of 1:0.7. A high pressure and high temperature Rockwell furnace was used to generate pressures up to 3.34 GPa and temperatures up to 1450°C. A 160 mg mass of $Mg_1B_{0.7}$ was put into a BN crucible, filling it to roughly 3/4 of its volume; the rest of crucible space was filled with BN powder. The crucible and a graphite heater were then placed in the middle of a pyrophyllite cube which was used as pressure medium for the high pressure furnace. The assembled cube is placed into the Rockwell furnace and pressure was applied at room temperature. After that the growth was (i) heated up to 1450°C over two hours, (ii) held at 1450°C for 1 hour, (iii) cooled down to 650°C over 6 hours, and (iv) cooled down to room temperature over 1 hour. Once cooled, the pressure was released. The pyrophyllite cube was then broken open and the BN crucible was removed and sealed in an evacuated amorphous silica tube for distillation of the excess Mg away from the hBN crystals. Distillation takes place over 3 hours, with one end of the sealed tube being held at 750°C and the other end hanging out of a clam furnace, around 100 – 200°C. After the distillation, the hBN crystals were mechanically separated from the BN crucible and $MgB_2$ single crystals.

### Device assembly

Exfoliation of few-layer flakes from the bulk $CrCl_3$ was carried out in an argon glove box to prevent hydration of the crystals. Optical contrast was employed to determine the thickness of each flake. The van der Waals magnetic tunnel junctions were then assembled by the dry transfer process in an argon environment.

The magnetic tunnel junctions were assembled by sequentially picking up flakes of boron nitride, few-layer graphite, $CrCl_3$, few-layer graphite, and boron nitride. The stacks were aligned such that the only vertical overlap between the two graphite electrodes is through the $CrCl_3$ tunnel barrier two to four layers in thickness. The use of top and bottom hexagonal boron nitride flakes provides protection from ambient conditions and an atomically flat substrate free of dangling bonds.

In each device, the final stack was transferred onto prepatterned Ti/Pd electrodes wirebonded to a chip carrier and loaded into a helium-3 cryostat for transport measurements.

**Transport measurements**

Magneto-transport measurements were carried out in a helium-3 cryostat with an external magnetic field applied either parallel or perpendicular to the device. Differential conductance measurements were obtained using low frequency lock-in methods (excitation frequency < 20 Hz). Our DC measurements were performed by applying a DC bias to the sample and reading out the DC current through a current preamplifier. All measurements were performed at fixed temperature (300 mK or 4.2 K), with the exception of the temperature-dependent data shown in Fig. 2c and 2d.

**Raman measurements**

Polarized Raman experiments were performed in a backscattering geometry using a confocal microscope spectrometer (Horiba LabRAM HR Evolution) with a 50x objective lens and 532 nm laser with a power of 2.0 mW. The spectrometer integration times were 5 minutes and 30 minutes for bulk and exfoliated crystals, respectively. Each scan was taken twice and then averaged before analysis.

The incident laser beam was linearly polarized in the vertical direction and a half-wave plate was placed just before the objective. The analyzer was placed in front of the spectrometer entrance and kept vertical for parallel configuration (XX). For polarization dependence, the half-wave plate was rotated at a step of 2.5º from 0º to 180º.

**DFT calculations**

Spin-polarized density functional theory (DFT) calculations were performed using PAW pseudopotentials[44] as implemented in the Vienna Ab initio Simulation Package (VASP)[45,46]. The energy cutoff for the plane wave basis was set at 350 eV with a 17x17x1 $\Gamma$-centered *k*-point grid. The Cl pseudopotential had 7 valence electrons ($3s^23p^5$) and Cr pseudopotentials with either 6 or 12 valence electrons were used ($3d^54s^1$ or $3p^63d^54s^1$, respectively). We used several exchange-correlation functionals: LDA[47], PBE[48], and optB86b-vdW exchange functional[49,50] with the vdW correlation functional[51]. Some calculations were performed using the LSDA+U method[52] with Hubbard U = 3 eV[29].

**Data Availability**: The data that support the findings of this study are available from the corresponding authors upon reasonable request.

# Supplementary Information for

# Giant enhancement of interlayer exchange in an ultrathin 2D magnet

Dahlia R. Klein[1†], David MacNeill[1†], Qian Song[2], Daniel T. Larson[3], Shiang Fang[3], Mingyu Xu[4,5], R. A. Ribeiro[4,5,6], P. C. Canfield[4,5], Efthimios Kaxiras[3,7], Riccardo Comin[1], Pablo Jarillo-Herrero[1*]

[1]Department of Physics, Massachusetts Institute of Technology, Cambridge, MA 02139, USA
[2]Department of Materials Science & Engineering, Massachusetts Institute of Technology, Cambridge, MA 02139, USA
[3]Department of Physics, Harvard University, Cambridge, MA 02138, USA
[4]Ames Laboratory, U. S. Department of Energy, Iowa State University, Ames, IA 50011, USA
[5]Department of Physics and Astronomy, Iowa State University, Ames, IA 50011, USA
[6]Centro de Ciências Naturais e Humanas, Universidade Federal do ABC, Santo André, SP, Brazil
[7]John A. Paulson School of Engineering and Applied Sciences, Harvard University, Cambridge, MA 02138, USA

[†]These authors contributed equally to this work.
[*]Correspondence to: pjarillo@mit.edu

**Contents:**

**I. Interlayer exchange model**

**II. Raman symmetry analysis**

**III. Density functional theory calculations**

**Fig. S1: Current-voltage curves**

**Fig. S2: Directional dependence of magnetoresistance**

**Fig. S3: Reversibility of magnetoresistance**

**Fig. S4: Additional trilayer device**

**Fig. S5: Temperature dependence of tetralayer device**

**Fig. S6: Raman spectra of bulk and exfoliated $CrCl_3$**

**Fig. S7: Polarization dependence of bulk crystal Raman peak**

**Fig. S8: Calculated energy difference versus layer spacing**

**Fig. S9: Calculated energy difference versus layer displacement**

## I. Interlayer exchange model

In order to extract the interlayer exchange value from the magnetoresistance curves, we employ a simple model to describe the magnetization energy in the few-layer $CrCl_3$ tunnel barrier. We consider two different energy contributions: the exchange energy between layers and the Zeeman energy from the external field. Here we assume that the out-of-plane magnetization is zero and can thus ignore out-of-plane shape anisotropy.

The interlayer exchange energy per unit area for a barrier of $n$ layers is:

$$U_E = \frac{1}{2} d\mu_0 M_s H_E \sum_{j=1}^{n-1} \hat{m}_j \cdot \hat{m}_{j+1}, \quad (S1)$$

where $d$ is the layer thickness, $M_s$ is the effective saturation magnetization, $H_E$ is the interlayer exchange field, and $\hat{m}_j$ is the unit magnetization vector of layer $j$. The Zeeman energy per unit area for an applied in-plane magnetic field $\vec{H}_{ext}$ is given by:

$$U_Z = -d\mu_0 M_s \vec{H}_{ext} \cdot \sum_{j=1}^{n} \hat{m}_j. \quad (S2)$$

By minimizing the total energy of the system $U = U_E + U_Z$, one can obtain a relationship between the external field and the exchange field as a function of magnetization directions in the layers. In the limit where the magnetizations all align with the external field, we obtain the value of the exchange field in terms of this saturation field required to fully polarize the layers.

We start with the bilayer $CrCl_3$ junction. We assume that the two layers' magnetization vectors $\hat{m}_1$ and $\hat{m}_2$ point equal and opposite from the applied magnetic field, illustrated below. Thus, the energy can be written and minimized with respect to a single parameter $\phi$:

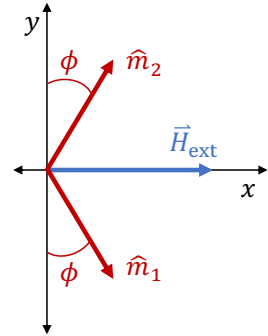

$$\frac{U}{d\mu_0 M_s} = \frac{1}{2} H_E \hat{m}_1 \cdot \hat{m}_2 - \vec{H}_{ext} \cdot (\hat{m}_1 + \hat{m}_2)$$

$$= -\frac{1}{2} H_E \cos(2\phi) - 2H_{ext} \sin \phi$$

$$\frac{\partial \left(\frac{U}{d\mu_0 M_s}\right)}{\partial \phi} = H_E \sin(2\phi) - 2H_{ext} \cos \phi = 0$$

When $H_{ext} \leq H_{sat}$, the above equation has a solution:

$H_{ext} = H_E \sin \phi$

This determines $\phi$ as a function of $H_{ext}$ in the antiferromagnetic state. The material is saturated when the two layers are fully aligned with the external field (i.e. $\phi = 90°$) such that $H_{ext}(\phi = 90°) = H_{sat}$ in. Thus, we find $H_{sat} = H_E$ for 2L $CrCl_3$.

For trilayer $CrCl_3$, we assume that the top and bottom layers have equal magnetization directions, but can have a different angle than the middle layer.

$$\frac{U}{d\mu_0 M_s} = \frac{1}{2}H_E(\hat{m}_1 \cdot \hat{m}_2 + \hat{m}_2 \cdot \hat{m}_3) - \vec{H}_{ext} \cdot (\hat{m}_1 + \hat{m}_2 + \hat{m}_3)$$
$$= -H_E \cos(\phi + \phi') - H_{ext}(2\sin\phi + \sin\phi')$$
$$\frac{\partial\left(\frac{U}{d\mu_0 M_s}\right)}{\partial \phi} = H_E \sin(\phi + \phi') - 2H_{ext}\cos\phi = 0$$
$$\frac{\partial\left(\frac{U}{d\mu_0 M_s}\right)}{\partial \phi'} = H_E \sin(\phi + \phi') - H_{ext}\cos\phi' = 0$$

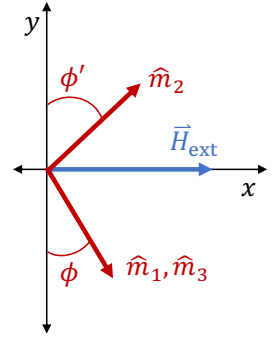

Using the equations above, we can derive the following relation:

$$\frac{H_{ext}}{H_E} = \frac{1}{2}\sin\phi' + \sqrt{1 - \frac{1}{4}(\cos\phi')^2}$$

We note that this relation only holds above a finite critical $H_{ext}$. This is because 3L $CrCl_3$ has a net moment and, for low $H_{ext}$, this moment aligns with the applied field direction while keeping the sublattice moments antiparallel ($\phi = 90°$, $\phi' = -90°$). When we fix the angle $\phi' = 90°$ such that the 3L $CrCl_3$ barrier is fully aligned with the external field, $H_{ext} = 3H_E/2$.

Finally, we consider four layers in which the outer two layers and inner two layers each have equal and opposite angles.

$$\frac{U}{d\mu_0 M_s} = \frac{1}{2}H_E(\hat{m}_1 \cdot \hat{m}_2 + \hat{m}_2 \cdot \hat{m}_3 + \hat{m}_3 \cdot \hat{m}_4)$$
$$-\vec{H}_{ext} \cdot (\hat{m}_1 + \hat{m}_2 + \hat{m}_3 + \hat{m}_4)$$
$$= -H_E \cos(\phi + \phi') - \frac{1}{2}H_E \cos(2\phi') - 2H_{ext}(\sin\phi + \sin\phi')$$
$$\frac{\partial\left(\frac{U}{d\mu_0 M_s}\right)}{\partial \phi} = H_E \sin(\phi + \phi') - 2H_{ext}\cos\phi = 0$$
$$\frac{\partial\left(\frac{U}{d\mu_0 M_s}\right)}{\partial \phi'} = H_E \sin(\phi + \phi') + H_E \sin(2\phi') - 2H_{ext}\cos\phi' = 0$$

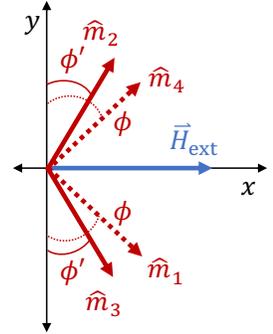

From the above equations, we obtain the following relation:

$$\frac{H_{ext}}{H_E} = \sin\phi' + \frac{1}{2}\sin\phi'\left(1 - \frac{H_E}{H_{ext}}\sin\phi'\right) + \frac{1}{2}\sqrt{1 - (\cos\phi')^2\left(1 - \frac{H_E}{H_{ext}}\sin\phi'\right)^2}$$

Evaluating the above expression for $\phi' = 90°$ gives a quadratic equation with solution $H_{ext} = (1 + \sqrt{2}/2)H_E$ for 4L $CrCl_3$.

## II. Raman symmetry analysis

CrCl$_3$ has two bulk crystallographic phases resulting in different symmetry-allowed Raman modes. At low temperature, it adopts a rhombohedral structure with space group $R\overline{3}$ and point group $C_{3i} = S_6$. At high temperature, it has a monoclinic structure with space group $C2/m$ and point group $C_{2h}$.

The phonon modes all form irreducible representations of the point group, yielding information about their symmetries and their Raman activity. The optical representations for the $C_{3i}$ and $C_{2h}$ point groups, respectively, are $4A_g + 3A_u + 4E_g + 3E_u$ and $6A_g + 4A_u + 6B_g + 5B_u$. The $A$ and $B$ representations are singlets; the $E$ representations are doublets.

The cross section (and thus intensity) for Raman scattering from mode $Q_i$ is proportional to:

$$I_i \propto \frac{d\sigma_i}{d\Omega} \propto \left| \hat{e}_s \frac{\partial \tilde{\alpha}}{\partial Q_i} \hat{e}_i \right|^2, \tag{S3}$$

where $\tilde{\alpha}$ is the polarizability tensor, $\frac{\partial \tilde{\alpha}}{\partial Q_i}$ is the Raman tensor for mode $Q_i$, and $\hat{e}_{i,s}$ are the polarization directions of the incident ($i$) and scattered ($s$) light.

Let us first consider the low temperature rhombohedral phase with point group $C_{3i}$. Raman active modes must be even under inversion, resulting in two Raman active irreps: $A_g$ and $E_g$. The Raman tensors are given by:

$$A_g = \begin{pmatrix} a & 0 & 0 \\ 0 & a & 0 \\ 0 & 0 & b \end{pmatrix}, \;\; ^1E_g = \begin{pmatrix} c & d & e \\ d & -c & f \\ e & f & 0 \end{pmatrix}, \;\; ^2E_g = \begin{pmatrix} d & -c & -f \\ -c & -d & e \\ -f & e & 0 \end{pmatrix} \tag{S4}$$

We assume a backscattering geometry with incident light along the -$z$ direction and reflected light along the +$z$ direction. For parallel polarization in the $xy$-plane making an angle $\theta$ with respect to the $x$-axis, we have $\hat{e}_i = \hat{e}_s = (\cos\theta, \sin\theta, 0)$. The Raman tensors are complex valued; we take the pairs $a$, $b$ and $c$, $d$ to have the same phase without loss of generality.

We analyze the polarization dependence of the 247 cm$^{-1}$ Raman peak in the main text (Fig. 4). At low temperatures in bulk CrCl$_3$, this corresponds to the doubly-degenerate $E_g$ mode. By plugging in the tensors from Eq. S4 into Eq. S3, we obtain:

$$I(^1E_g) \propto \left| \hat{e}_s\,^1E_g\,\hat{e}_i \right|^2 = |c\cos(2\theta) + d\sin(2\theta)|^2$$
$$= c^2 (\cos(2\theta))^2 + 2cd \cos(2\theta)\sin(2\theta) + d^2 (\sin(2\theta))^2$$
$$I(^2E_g) \propto \left| \hat{e}_s\,^2E_g\,\hat{e}_i \right|^2 = |d\cos(2\theta) + c\sin(2\theta)|^2$$
$$= d^2 (\cos(2\theta))^2 - 2cd \cos(2\theta)\sin(2\theta) + c^2 (\sin(2\theta))^2$$

We can see that the two components of intensity vary oppositely with polarization angle. However, since the $E_g$ mode is degenerate, these contribute to a peak at the same frequency. Summing the two intensities gives a constant $c^2 + d^2$. Thus, the 247 cm$^{-1}$ peak in the rhombohedral structure should be polarization independent, consistent with our observations of the bulk peak energy at 80 K (Fig. 4d).

In contrast, let us consider the high temperature monoclinic phase with point group $C_{2h}$. Its two Raman active irreps are $A_g$ and $B_g$ with the following Raman tensors:

$$A_g = \begin{pmatrix} a & 0 & d \\ 0 & b & 0 \\ d & 0 & c \end{pmatrix}, \; B_g = \begin{pmatrix} 0 & e & 0 \\ e & 0 & f \\ 0 & f & 0 \end{pmatrix} \tag{S5}$$

The high temperature 247 cm$^{-1}$ peak reflects the two nearly-overlapping $A_g$ and $B_g$ modes at slightly different frequencies. The polarization angle dependence of these two modes is again obtained by substituting Eq. S5 into Eq. S3:

$$I(A_g) \propto |\hat{e}_s A_g \hat{e}_i|^2 = |a(\cos\theta)^2 + b(\sin\theta)^2|^2$$
$$= \frac{1}{4}(a^2+b^2) + \frac{1}{8}(a+b)^2 + \frac{1}{8}(a-b)^2\cos(4\theta) + \frac{1}{2}(a^2-b^2)\cos(2\theta)$$
$$I(B_g) \propto |\hat{e}_s B_g \hat{e}_i|^2 = |e\cos\theta\sin\theta + e\cos\theta\sin\theta|^2$$
$$= e^2(\sin(2\theta))^2$$

Since the $A_g$ and $B_g$ modes in the $C2/m$ phase pair to form the degenerate $E_g$ mode in the $R\bar{3}$ phase, the parameters in the Raman tensors for these sets of modes must have matching symmetry patterns. Therefore, the 247 cm$^{-1}$ peak's $A_g$ mode has $a \sim -b$, resulting in an intensity $I(A_g) \propto a^2(\cos(2\theta))^2$. The observed Raman signal will be the sum of intensities from the two modes at different frequencies, which have opposite 4-fold dependences on the polarization angle. Thus, the overall peak should oscillate in energy in a 4-fold pattern versus polarization angle. This modulation is observed in the bulk crystal Raman signal at 300 K (Fig. 4d), as well as the exfoliated thin crystal Raman signals at both 300 K and 80 K (Fig. 4e).

We note that while the Raman tensor components can be complex, the relative phases between the entries cancel out in the intensities of the $B_g$ and $E_g$ modes. For the $A_g$ mode, a small relative phase between $a$ and $b$ would give a small constant contribution to the intensity independent of polarization angle. This correction does not affect the conclusions of the symmetry analysis.

### III. Density functional theory calculations

We performed spin-polarized density functional theory (DFT) calculations of bilayer CrCl$_3$ with magnetic moments of the Cr atoms aligned within each layer. We calculate the total energy of the bilayer system assuming the magnetic moments of the two layers are either aligned (FM) or opposite each other (AFM). The energy difference, defined as $\Delta E = E(\text{FM}) - E(\text{AFM})$ below, can then be determined for different stacking arrangements of the layers. We use PAW pseudopotentials[1] as implemented in the Vienna Ab initio Simulation Package (VASP)[2,3]. For the calculations to converge, the energy cutoff for plane wave basis is set at 350 eV with a 17x17x1 Brillouin zone sampling in $k$-space. Because calculations of magnetic energy differences are very sensitive to the DFT setup, we have repeated the calculations under various conditions. We employed a Cl pseudopotential with 7 valence electrons (3s$^2$3p$^5$) and considered Cr pseudopotentials with either 6 or 12 valence electrons (3d$^5$4s$^1$ or 3p$^6$3d$^5$4s$^1$, respectively). Aside from the conventional LDA[4] and PBE[5] functionals, we also considered the optB86b-vdW exchange functional[6,7] with the vdW correlation functional[8]. In addition, we considered the effect of the on-site Coulomb interaction for the Cr d states using the LSDA+U method[9] with Hubbard U = 3 eV[10]. We have chosen nine representative combinations of pseudo-potentials, exchange-correlation functionals, and U values to demonstrate the overall dependence of magnetism on stacking order.

First, we consider the energy difference between the FM and AFM states versus layer spacing for interlayer alignments of bilayer $CrCl_3$ corresponding to the bulk $C2/m$ and $R\bar{3}$ phases (Fig. S8a and S8b). The $C2/m$-type plots (Fig. S8a) have a positive energy difference between the FM and AFM states, indicating a strong preference for AFM coupling in the ground state of bilayer $CrCl_3$. In contrast, the $R\bar{3}$-type curves (Fig. S8b) vary between positive and negative energy differences depending on the chosen DFT functional and layer separation, therefore indicating a weaker preference for AFM coupling. Nonetheless, the difference between the two phases (Fig. S8c) shows that the $C2/m$-type phase always has a much larger AFM coupling strength than the $R\bar{3}$-type phase regardless of choice of DFT functional.

Second, we generalize our analysis beyond the $C2/m$-type and $R\bar{3}$-type configurations to examine the overall dependence of interlayer coupling on relative stacking at fixed layer separation (Fig. S9). At a realistic interlayer spacing value of 6 Å, we find that $R\bar{3}$-type stacking sits at a local maximum favoring FM coupling, while $C2/m$-type stacking lies near the local minimum favoring AFM coupling.

Together, these DFT calculations are in agreement with our observed results from tunneling magnetoresistance and Raman spectroscopy measurements. They support the conclusion that ultrathin $CrCl_3$ adopts a $C2/m$-type stacking order at low temperatures and therefore exhibits markedly stronger AFM interlayer coupling than bulk $CrCl_3$ in the $R\bar{3}$ configuration.

As a final quantitative check, we compare the energy difference in Fig. S8c with estimated energy differences for ultrathin and bulk $CrCl_3$ from experimental data. The work per unit volume required for an applied magnetic field to magnetize a material is given by $dW = \mu_0 H dM$. During the magnetization process of bilayer $CrCl_3$, the component of $M$ along the applied field direction obeys $M(H) = M_s H/H_{sat}$ (see Section I). Therefore, to fully saturate $CrCl_3$, the total work density is $W = \int_0^{M_s} dM \mu_0 H_{sat} M/M_s = \mu_0 H_{sat} M_s/2$. Since $M_s = 3\mu_B$ per Cr atom volume, the work done per Cr atom is $W = 3\mu_0 \mu_B H_{sat}/2$. As the AFM to FM transition in $CrCl_3$ occurs via a coherent and reversible rotation of the moments (see Fig. S3), this is also the (free) energy difference between the AFM state and the fully polarized FM state $\Delta E = E(FM) - E(AFM)$. To estimate the value for a $C2/m$-type bilayer, we use the observed saturation field $\mu_0 H_{sat}$ of 0.86 Tesla for bilayer $CrCl_3$. We multiply our value by 4 to match the 4 Cr atoms used in the DFT calculation, yielding an estimate for the energy difference $\Delta E = 0.30$ meV. Next, we estimate the $R\bar{3}$-type bilayer energy difference. We assume that the saturation field in and $R\bar{3}$-type bilayer would be half of the bulk $R\bar{3}$ saturation field (0.25 Tesla) since each layer only couples to one adjacent layer rather than two. Therefore, we find that the energy difference is $\Delta E = 0.043$ meV. Finally, we subtract the two values for $\Delta E$, yielding a difference in the interlayer exchange per unit cell between the two stacking types of approximately 0.26 meV. This value is on the same order as the energy difference in Fig. S8c for realistic values of layer separation from 0.55 nm to 0.65 nm.

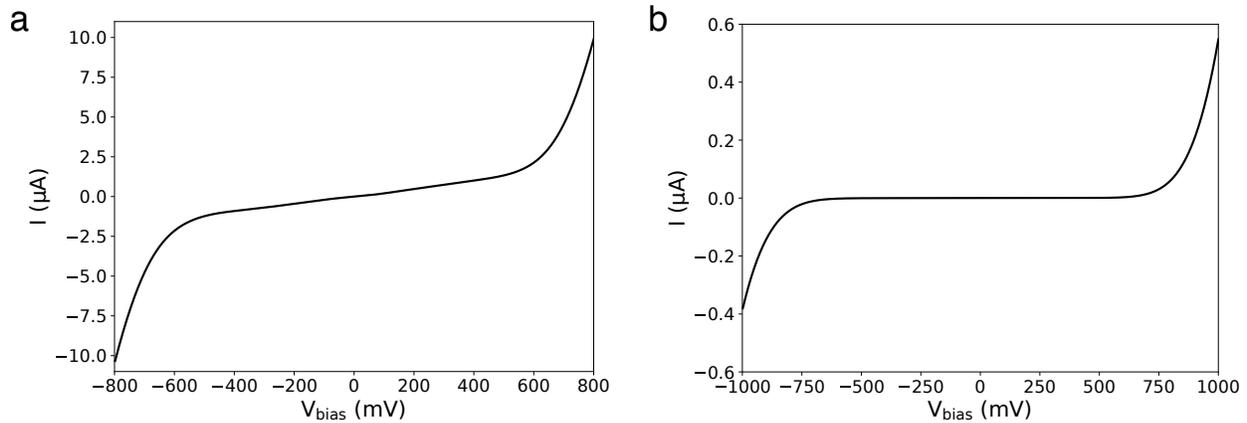

**Figure S1 | Current-voltage curves.** Current versus bias voltage for **(a)** bilayer and **(b)** tetralayer $CrCl_3$ devices at 4.2 K with an AC excitation of 50 mV. The DC current values were calculated by integrating the differential conductance and fixing the current to be zero at zero applied DC bias. The sharp increase of current at finite bias indicates tunneling in the Fowler-Nordheim regime.

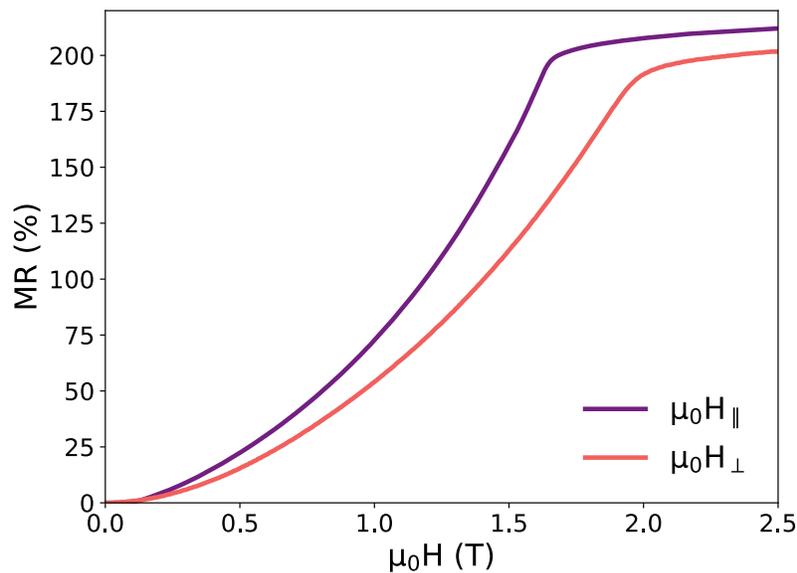

**Figure S2 | Directional dependence of magnetoresistance.** Magnetoresistance versus applied magnetic fields in the in-plane (purple) and out-of-plane (pink) directions for a tetralayer $CrCl_3$ device. The out-of-plane magnetoresistance saturates at a field of 1.95 Tesla, 0.3 Tesla larger than the in-plane saturation field of 1.65 Tesla. This difference is consistent with the observed demagnetization field of 0.35 Tesla in bulk $CrCl_3$[11].

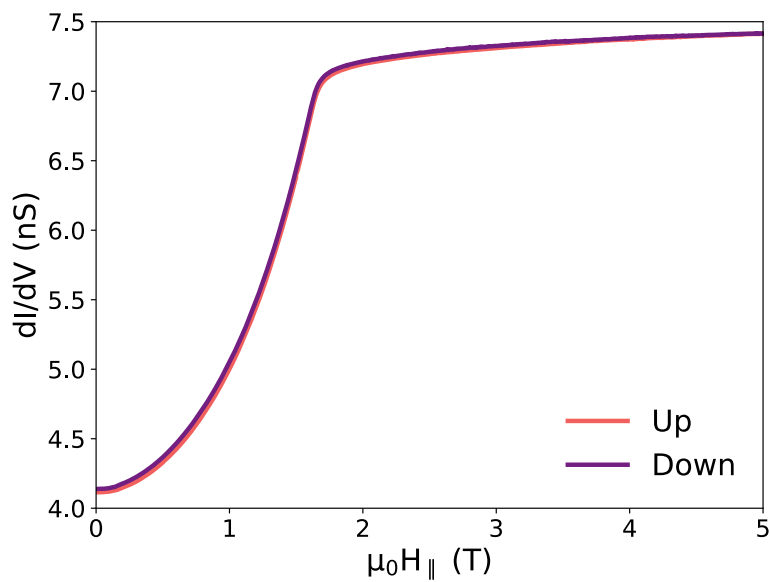

**Figure S3 | Reversibility of magnetoresistance.** Differential conductance versus applied in-plane magnetic field swept up (pink) and down (purple) for a tetralayer CrCl$_3$ device at 4.2 K with an AC excitation of 500 mV. The traces demonstrate no hysteresis in the tunnel junction magnetoresistance.

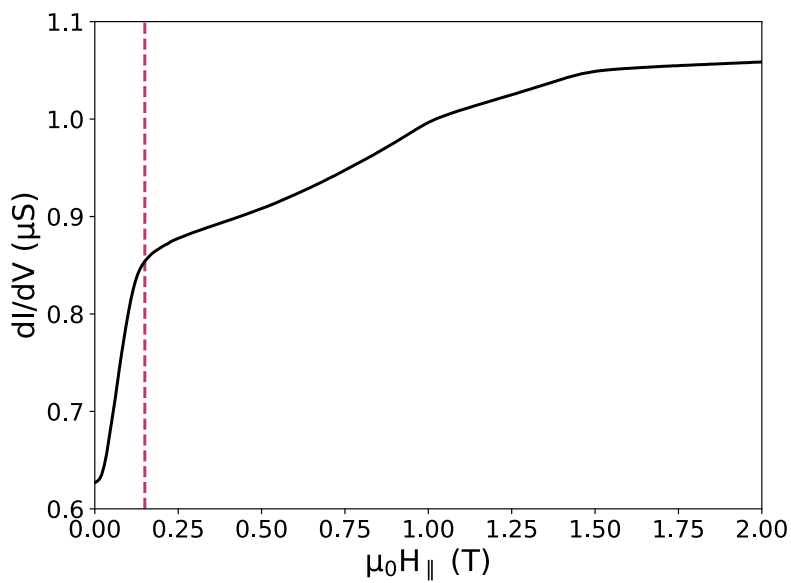

**Figure S4 | Additional trilayer device.** Differential conductance versus applied in-plane magnetic field for a different trilayer CrCl$_3$ tunneling device taken at 4 K with a bias voltage of -625 mV and AC excitation of 50 mV. The dashed purple line denotes an applied field of 0.15 T. Using this value for the saturation field for 3L CrCl$_3$ gives an interlayer exchange field of approximately 0.1 T, similar to the reported bulk value of 0.084 T[12]. This observation suggests that the CrCl$_3$ barrier in this device may have domains of both rhombohedral and monoclinic stacking.

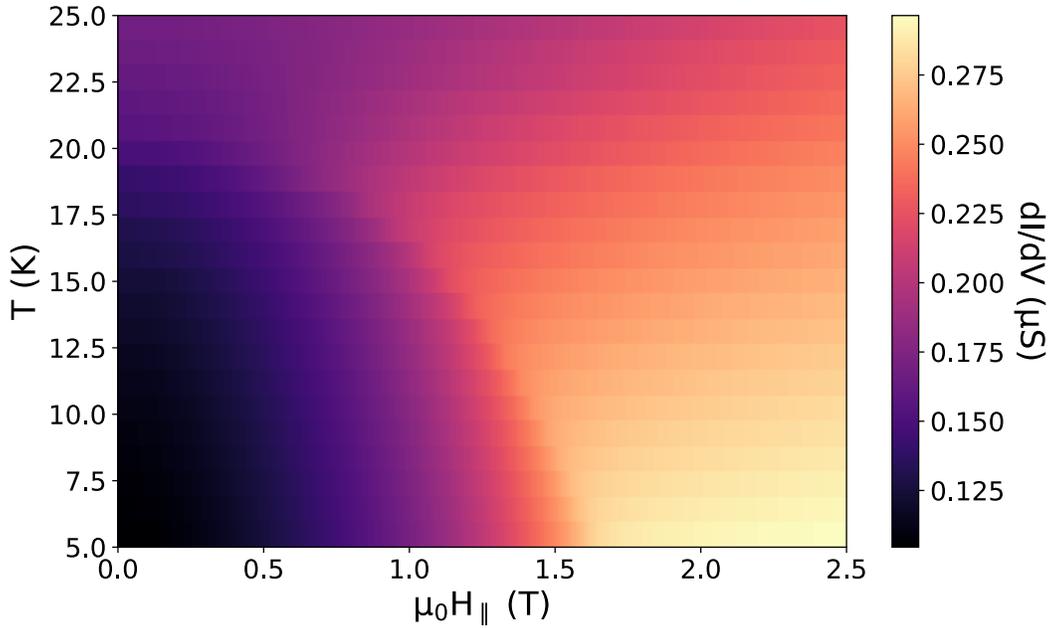

**Figure S5 | Temperature dependence of tetralayer device.** Differential conductance versus applied in-plane magnetic field and temperature. The DC bias is 750 mV and the AC excitation is 50 mV.

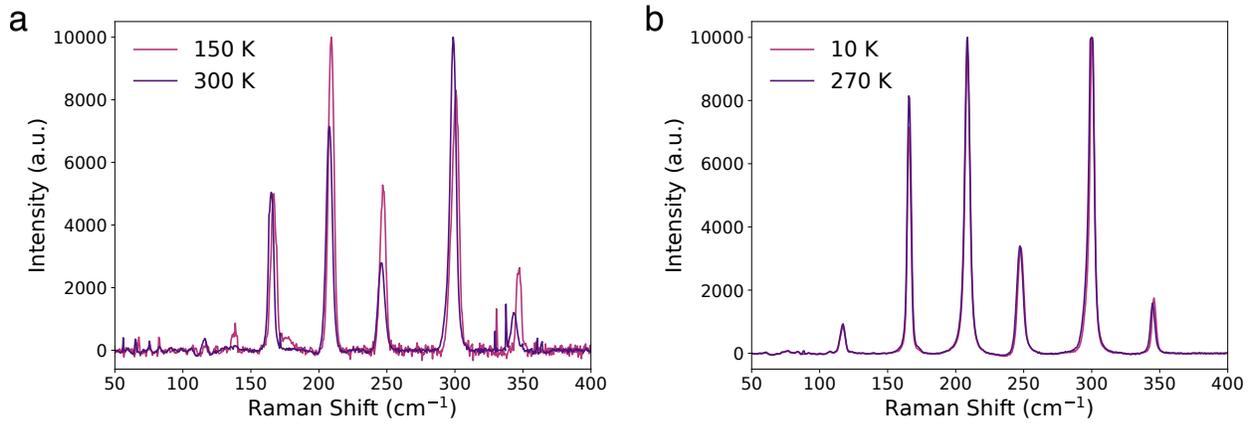

**Figure S6 | Raman spectra of bulk and exfoliated CrCl$_3$.** Raman spectra at high and low temperatures of **(a)** a bulk crystal and **(b)** an exfoliated 35 nm thick flake. The spectra were taken using a 532 nm laser and parallel configuration (XX). The background has been subtracted from each trace.

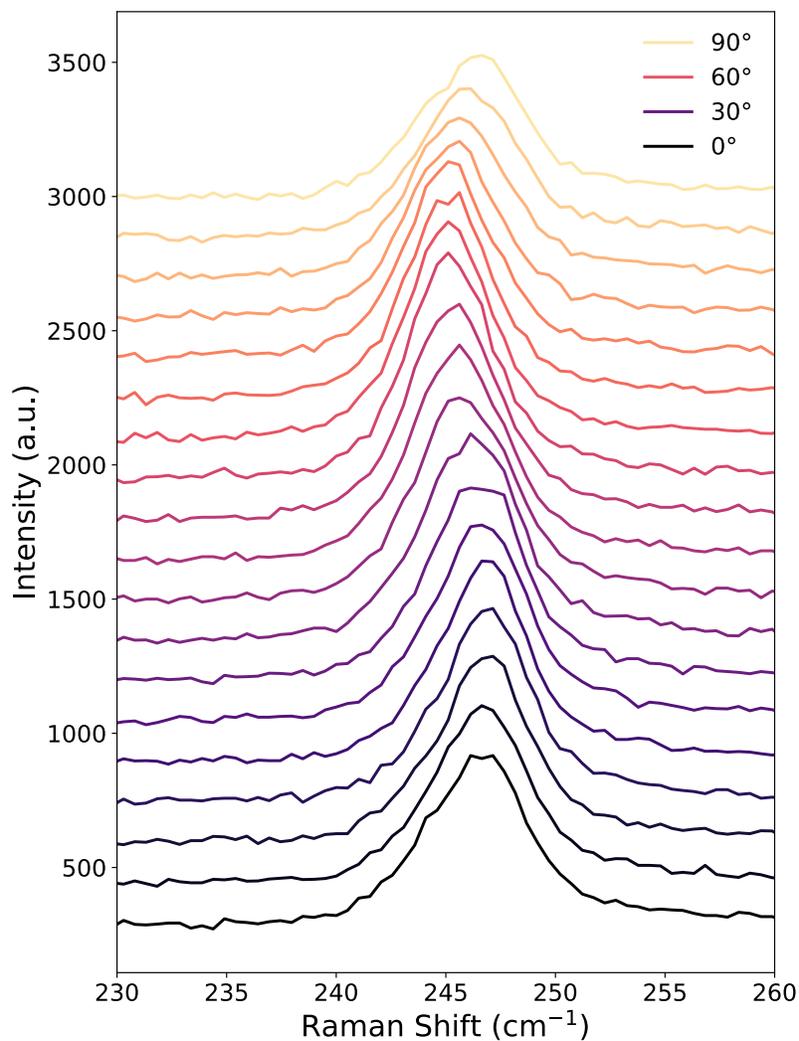

**Figure S7 | Polarization dependence of bulk crystal Raman peak.** Raman spectra of the 247 cm$^{-1}$ peak in a bulk crystal with polarizations ranging from 0º to 90º in 5º steps. The data were taken at 300 K and the polarization was changed using a half-wave plate rotated from 0º to 45º in 2.5º steps.

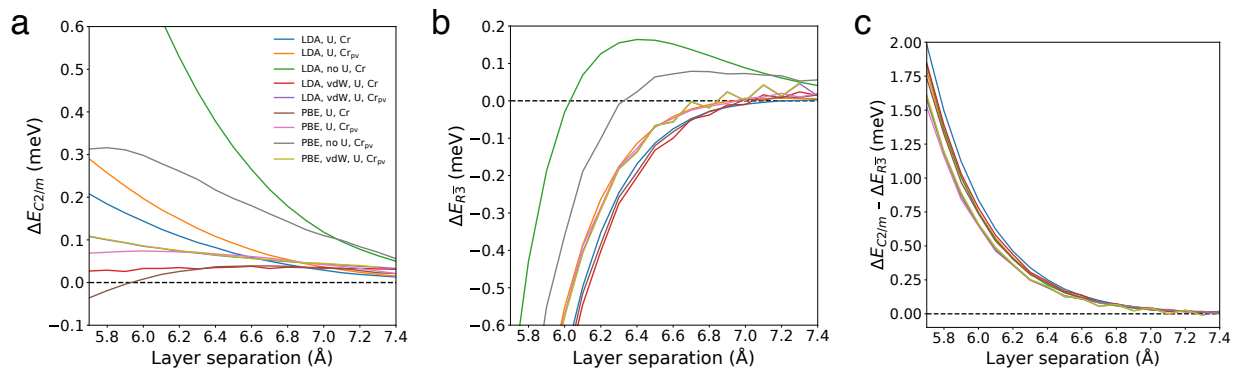

**Figure S8 | Calculated energy difference versus layer spacing.** Calculated energy difference between FM and AFM states for bilayer CrCl$_3$ as a function of interlayer separation in the **(a)** $C2/m$-type and **(b)** $R\bar{3}$-type configurations using different DFT functionals. **(c)** shows the difference between the $C2/m$-type and $R\bar{3}$-type energy differences versus interlayer separation.

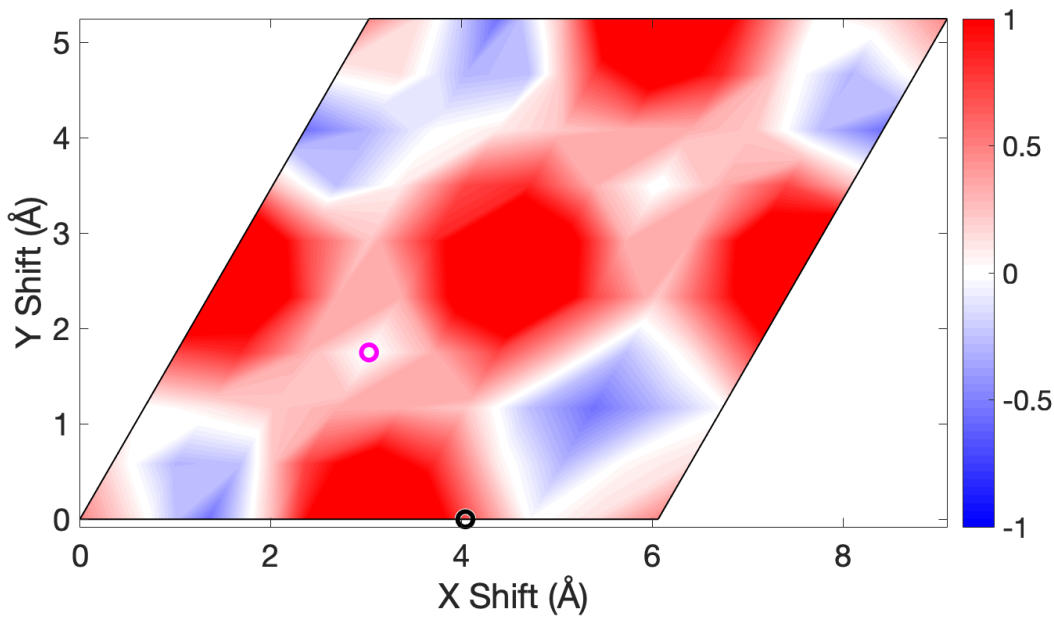

**Figure S9 | Calculated energy difference versus layer displacement.** Calculated energy difference (in meV) between FM and AFM states for bilayer CrCl$_3$ as a function of interlayer displacements in the in-plane X and Y directions. The layer separation is fixed at 6 Å and the DFT functional used is LDA using a Cr pseudo-potential. The $C2/m$-type and $R\bar{3}$-type stacking phases are indicated by the black and magenta circles, respectively.